# Diversity Assessment Based on a Higher Similarity-Higher Entropy Relation after Rejection of Gibbs Paradox

Shu-Kun Lin

Molecular Diversity Preservation International (MDPI), Saengergasse 25, CH-4054 Basel, Switzerland

Tel: +41 79 3223379, Fax +41 61 3028918

E-mail: Lin@mdpi.org, URL: http://www.mdpi.org/lin.htm

21 October 1999

**Abstract**. The diversity of the symbols of the information source is calculated following the definition that entropy is the information loss and following a new entropy-symbol similarity relation after the rejection of the Gibbs paradox statement. Diversity in a range of 0 to 1, an index similar to Shannon's redundancy, decreases with the increase in the species similarities. A pairwise similarity formula has been defined and used to demonstrated that the diversity expression gives the expected diversity. The higher entropy-higher similarity relation leads to the higher information-higher diversity relation.

**Keywords:** Species diversity, entropy, information loss, similarity, diversity calculation.

## 1. Introduction

Biodiversity assessment has been a hot topic of practical importance for preservation of existing biological species in a rational way. With the development of high throughput screening technology (a very fast way of testing if a molecule is active as a drug) in the pharmaceutical industry in recent years, the acquisition of molecular samples by collection and combinatorial or parallel, automatic synthesis has now become the bottleneck in the process of new drug discovery. The molecular diversity assessment also became an urgent topic [1,2].

There are many methods of calculating diversity of species (either biological or chemical species). Natually the best approach should be the information theoretical method, using information theoretical concepts [3]. We tried to set up a new method of diversity calculation [1]. It is based on a clear definition that entropy is the information loss [4] and a new relation of entropy-similarity (Figure 1c) [5-7] constructed after rejection of the Gibbs paradox statement (Figure 1a) [8]. In information theory discussed in communication, all the symbols used for encoding are assumed to be distinguishable. However, we discuss the species diversity of these symbols (individual molecules in molecular diversity and animals in biodiversity considerations). Therefore, it is the similarity of these symbols and their relation to entropy that are of particular interest to us, and we must adapt the formulas used by Shannon to the situation concerned here. Furthermore, the consideration of the diversity and the similarity of these "symbols" (molecules or other individules) at the "information source" (the considered system) are very significant in the studies of the structural stabilities in physics, chemistry and biology.

In this paper, simple examples and a pairwise similarity formula suitable for assessing molecular diversity has been presented and used for species diversity calculation.

## 2. Entropy-information loss and entropy-similarity

The choice of the definition that entropy is information loss [4] is convenient because it is at least already widely accepted by biologists and chemists. Entropy is defined here as a concept equivalent to Shannon's uncertainty.

Another reason of using "entropy is information loss" definition is that there are only two kinds of information losses: one is due to dynamic motion which is illustrated in Figure 2; the other is due to intrinsic similarity of the symbols or the physical or conceptual entities (biological or chemical species, or 1 and 0 used for a binary system). Both information losses can happen actually in examples of communication (see the following).

Shannon defined entropy ($H$) as a measures of information, choice and uncertainty. Entropy, denoted here as $S$, is given by the familiar expression (the positive coefficient is taken as 1):

$$S = -\sum_{i=1}^{w} p_i \ln p_i \tag{1}$$

where $p_i$ is the probability of the $i$th microstate with the property that

$$\sum_{i=1}^{w} p_i = 1 \tag{2}$$

Hrere we use a microstate as a symbol defined in Shannon's information source (Sometimes we will also use microstate as a message, which is made out of symbols. A microstate can be regarded as a composite symbols. An example of the composite symbol is the edcoding of A, B, C, D, etc by binary symbols 0 and 1. An "A" can be represented by a long composite symbol, a longer sequence of 0 and 1). Information loss due to dynamic motion is simple. If a set of all possible symbols all appears simultaneously with the identical probabilities, the situation is most chaotic (figure 2). Consequently there is no actual message selected. This is due to the *similarity* of the probability values of all the possible messages. If these probability values are the most similar (or the same), we have the highest entropy ($S=S_{max} = \ln w$) following the well-known inequality

$$-\sum_{i=1}^{w}(p_i \ln p_i) \leq \ln w \tag{3}$$

Then, the information has the minimum value (zero, *I*=0, see the following section).

Even though it has not been discussed in Shannon's classical paper [3], the information loss due to the intrinsic similarity of symbols (and consider if these symbols belong to a set of distinguishable species by comparison, or consider directly the similarity of the species themselves) at the information source seems also clear. If all the fonts used in "I love her" are extremely similar (or the same) and this font is represented by @, we will have a sentence reads "@ @@@@ @@@" (Figure 3). If a typewriter has all keys actually extremely similar, to say all actually like "@". A discrete source using these extremely similar *w* microstates (symbols) actually cannot generate any information. If you send a telegraph by using this set of 26 symbols instead of the distinguishable 26-symbol "alphabet" (the 26 letters), of course the information loss will be obvious. Actually these 26 symbols are all @. Applying quations 1 and 2 will give the maximum entropy (Expression 3). Therefore, both cases satisfy the general relation of higher similarity-higher entropy (or higher information loss) (Figure 1c).

Incidentally it should be pointed out that the higher entropy-higher similarity relation Figure 1c) holds universally true, where the mathematical proof is simply rooted at the well-known inequality (3) where the right side is the maximum entropy because all the *w* microstates or symbols are extremely similar or the same. This is summarized as the similarity principle: *If all the other conditions remain constant, the higher the similarity among the components is, the higher value of* *entropy of the mixing (for fluid phases), the assembling (for solid phases) or any other analogous processes (of assemblage formation, such as quantum states in quantum mechanics) will be, the more spontaneous the processes will be, and the more stable the mixture and the assemblage will be.* Practically this conforms to and explained all the related experimental facts, e.g., phase separation. Different substances do not mix but spontaneously separate because the indistinguishable substances are the most spontaneously miscible ones. In other words, as a consequence of the most spontaneously mixing of the most similar (indistinguishable) substances, different substances separate. Theoretically this is also conforms to, and can be used to prove Curie's symmetry principle (the effects are more symmetric than the causes) following a "higher similarity-higher symmetry-higher entropy-higher stability" relation. After the proof of the symmetry principle, and after the the connection of the symmetry principle and the second law being established, many outstanding phase transition problems (symmetry breaking probems) can be elegantly solved also [5,6].

It should be emphasized that a clear definition that entropy is information loss should be both very convenient and very necessary for further development in information theory towards its application in structural stability and process spontaneity characterization in physics, chemistry and biology.

## 3. The formula

We define diversity on a relative scale [1]. The diversity index ($D$) defined here is equivalent to Shannon's definition of redundancy. It is defined as the ratio of the information ($I$) and the maximum information ($I_{max}$),

$$D = \frac{I}{I_{max}} \tag{4}$$

Entropy is unambiguously defined as information loss by the following relation:

$$I = S_{max} - S \tag{5}$$

In this equation, entropy is given by the familiar expression (euqtion 1), where $p_i$ is the probability of the $i$th microstate with the property that

$$\sum_{i=1}^{w} p_i = 1$$

while the maximum entropy is $S_{max} = \ln w$, where $w$ is the indistinguishability number which is the number of microstates (or symbols) of indistinguishable property [1]. Because both information and entropy are logarethmic functions, both are never negative. Their minimum values are both zero. Their maximum values are the same value also ($I_{max} = S_{max} = \ln w$).

The apparent indistinguishability number of microstates (or symbols) is defined as

$$w_a = \exp(-\sum_{i=1}^{w} p_i \ln p_i) \qquad (6)$$

and equation 1 becomes

$$S = \ln w_a \qquad (7)$$

which is the logarithmic relation of entropy and indistinguishability. Now the number of the indistinguishable symbols is $w_a$.

Practically, in order to record information at the information source, a system composed of *N* "unit devices" is used. In computer science or in our daily information recording as well, these "unit devices" are *N* symbols assembled on a media such as a piece of paper. These symbols appear as *M* attributes, based upon which it is said that the system has *M* species, such as the two species 0 and 1 in the binary system [3].

Because the assessment of diversity of *N* symbols (molecules or chemical samples for molecular diversity and a plant or an animal for biodiversity) is our only concern [4], the number of symbols (*N*) and the maximum species number (*M*) are designated as the same: $N = M$. This can be envisaged as *N* holes in microplates used for bioactivity screening containing *N* compound samples. Normally *N* can be either greater or smaller than *M*. For example, a harddisk of *N* bits has *N* symbols with *M* equals 2. Normally if one puts 200 Chinese characters in a typical page of paper written in Chinese where *N* equals 200 and (here the species number *M* is the total number of different Chinese characters normally used which is 10000).

If these *N* symbols are all distinguishable, they can be used to record the maximum information as given by equation 8.

$$I_{max}(N, N) = \ln w = \ln N^N = N \ln N \qquad (8)$$

For instance, if red ink is used to represent 0 and blue ink 1, and two bottles of these different inks are used, 2 bits of information can be recorded if the number of symbol ($N$) is 2. There will be 4 ($w = 2^2 = 4$) *distinguishable* microstates, see Figure 4). It is said that this is the maximum information (equation 8) because one can still intentionally use only a small part of the available species to record only smaller amount of information. In equation 8, $w$ is the number of *distinguishable* microstates:

$$w = N^N \tag{9}$$

From equation 5, the corresponding entropy has the minimum value which is zero:

$$S_{min}(N,N) = \ln 1 = 0 \tag{10}$$

This extreme case is illustrated in Figure 4 ($N=2$, $w=4$) and Figure 5 ($N=3$, $w=27$) [8].

Let all the $N$ samples in the $N$ bottles are the samples of extremely similar (or the same) property. Then there will be still $w = N^N$ microstates (or $N^N$ assemblages) constructed by the $N^N$ times of different combinatorial sequences of assembling to form solid structures. However, because they are all virtually *indistinguishable* microstates, there is always the minimum information and the maximum entropy (eqs 11 and 12):

$$I_{min}(N,N) = \ln 1 = 0 \tag{11}$$

$$S_{max}(N,N) = \ln w = \ln N^N = N \ln N \tag{12}$$

For example, suppose you have accidentally installed two bottles of red ink for a printer. Even though exactly the same amount of effort is taken to prepare the 4 microstates, i.e., the four microstates are prepared in a same way as that of Figure 4 by using inks from the two individual bottles, there will be 4 ($w = 2^2 = 4$) indistinguishable microstates (see Figure 6). Similarly, whether we factually take the same sample from one sample bottle or different sample bottles in bioactivity test laboratory, we always have $w$ *indistinguishable* microstates, if they are virtually the same compound in all the $N$ bottles; see Figure 6 ($N=2$, $w=4$, if all species are factually 0) and Figure 7 ($N=3$, $w=27$, if all species are B). The maximum microstate indistinguishability number is therefore

$$w = N^N \tag{13}$$

These two extreme sets of distinguishable and indistinguishable microstates, which give the minimum (zero) entropy and the maximum entropy values, respectively (equations 10 and 12), already illustrated our unique approach based on our similarity-entropy relation (see also Figure 1c).

Generally, suppose the $N$ symbols used to construct $N^N$ microstates are only mutually similar to certain extent among them and they are neither distinguishable (equations 8 and 10) nor indistinguishable (equations 11 and 12). Instead of using equation 1 directly, eq 14 is used to calculate entropy.

$$S(N,N) = -\sum_{j=1}^{N}\sum_{i=1}^{N}(p_{ij}\ln p_{ij}) \qquad (14)$$

The $N^2$ pairwise similarities $r_{ij}$ in the table

$$\begin{array}{ccccc} r_{11} & r_{12} & \cdots & \cdots & r_{1N} \\ r_{21} & r_{22} & \cdots & \cdots & r_{2N} \\ \cdots & \cdots & \cdots & \cdots & \cdots \\ \cdots & \cdots & \cdots & \cdots & \cdots \\ r_{N1} & r_{N2} & \cdots & \cdots & r_{NN} \end{array} \qquad (15)$$

have values limited between 0 and 1 and are given by pairwise comparison among the $N$ symbols according to one and only one systematically followed standard of comparison for all the values $p_{ij}$ ($i=1,\cdots,N; j=1,\cdots,N$). A normalization factor $c$ is required also:

$$c = \frac{1}{\sum_{i=1}^{N} r_{ij}} \qquad (16)$$

It follows that

$$p_{ij} = c r_{ij} \qquad (17)$$

Then equation 14 will give the same results as given by eqs 10 and 12 directly and respectively under the two extreme conditions.

In principle, the general equation (equation 4) should be directly used, where $w$ is simply replaced by $N^N$. The obvious disadvantage of using equation 4 directly is that the sum runs all over the $N^N$ microstates (see Figures 4-7). The calculation of these terms of enormous number $N^N$, which can be an astronomical figure, is impractical. Normally $N$ is 100000, the size of a compound sample library or sublibrary. In equation 14, the number is substantially reduced to totally $N^2$ terms of $p_{ij} \ln p_{ij}$.

Secondly, we may not be really interested in using the chemical samples or individual molecule to record information by taking the sample bottles as "unit devices". Therefore, we will not perform experiment or calculation to characterize the chemical structural and other physicochemical properties of all these $N^N$ *microstates* (combinatorial assemblings). Instead, we measure (or calculate from the known structures) the properties of the $N$ *individuals*, based on which the pairwise similarities between any two compounds are to be relatively much easily calculated. This means that, instead of considering similarities and probabilities among $N^N$ *microstates*, only $N^2$ probabilities $p_{AA}$, $p_{AB}$, $p_{AC}$, etc., calculated from the pairwise similarities among the $N$ *individuals* A, B, C, ..., etc., will be considered.

The first column of Table 1 showed several sets of three imaginary compound samples or symbols (A, B and C) on a uniform property scale as used by Agrafiotis [10]. If the properties of the samples are the same, the points will coincide and these three samples all will be regarded as the same samples. If their distances are very short and they are very close, they are regarded as very similar.

The probability $p_{ij}$ calculated from equation 17 means the probability of finding the *j*th symbol as the *i*th species, or the *j*th sample as the *i*th chemical species.

The diversity of these symbols is yet to be assessed and unknown; they are presumably similar to each other to certain extent among them [1]. Therefore, the comparisons are *not* performed between the $N$ individuals and a set of *a priori* known set of *distinguishable* prototypes; the comparisons are performed among the $N$ individuals themselves directly. The normalization factor $c$ is required because these values are subject to the constraint:

$$\sum_{i=1}^{N} p_{ij} = 1 \tag{18}$$

Using of the logarithmic expression of entropy

$$S = \ln w_a = N \ln s_a \tag{19}$$

we define the apparent species indistinguishability number $s_a$. For the examples shown in Figures 6 and 7, $s_a = N$. Generally,

$$s_a = \sqrt[N]{e^S} \tag{20}$$

Easily, the apparent species number $M_a$ can be calculated.

$$M_a = \exp\left(\ln N + \frac{1}{N}\sum_{j=1}^{N}\sum_{i=1}^{N} p_{ij} \ln p_{ij}\right)$$
$$= \frac{N}{s_a} \tag{21}$$

Using this method properly, the molecular diversity as expressed by the diversity index $D$ and several related parameters can be calculated. For instance, to compare the diversity of several selections of a sublibrary of compounds (molecules) from all available chemical sources, and to acquire the same number of samples of the highest diversity for many different screening purposes, the sublibrary of minimized entropy $S$ and the maximized information $I$ is desirable.

## 4. Pairwise Similarity Definition and Calculation

Before calculating the biodiversity or molecular diversity of $N$ individuals (symbols), the similarities for all the mutual pairwise comparisons among all the $N$ individuals should be clearly defined. Whether it is a proper definition of pairwise similarities can be quickly checked first by the following criteria of the two extreme cases: (a) If $N$ symbols are all distinguishable, the entropy of this system is the minimum which is zero (equation 10). (b) If $N$ species are all indistinguishable, the entropy of this system is the maximum (equation 12).

If the pairwise similarities are estimated directly by either

$$r_{ij} = 1 - d_{ij} \tag{22}$$

or

$$r_{ij} = \frac{1}{1+d_{ij}} \tag{23}$$

which might be suitable for other methods [10], the maximum number of distinguishable species is only 2, located at the two ends of the property scale, 0.00 and 1.00, respectively. The distance $d_{ij}$

10between these two least similar species is 1.00. This does not conform with the first simple criterion (equation 10).

Note, equation (23) does not conform with the second criterion either: The minimum similarity value is 0.5, instead of zero, which is normally the minimum value of a properly defined similarity scale [11]. The minimum value corresponds to the largest distance, which is 1 in Agrafiotis' definition [10].

We propose that, instead of equation 22, the following formula of pairwise similarity, which clearly conforms with the two simple criteria, is adopted:

$$r_{ij} = \begin{cases} 0, & \text{if } d_{ij} \geq \frac{1}{N-1} \\ 1 - d_{ij}(N-1), & \text{if } d_{ij} < \frac{1}{N-1} \end{cases} \quad (24)$$

In this formula, the shortest distance, which defines that certain two species are distinguishable, is

$$d_{ij} = \frac{1}{N-1} \quad (25)$$

provided that the property scale range is [0, 1]. Again, remember that "distinguishability" means the least similarity. If the distance is shorter than $1/(N-1)$ (equation 25), the two considered samples are similar. If they coincide, they are indistinguishable samples or the same samples. If $d_{ij}$ is no less than $1/(N-1)$, $r_{ij}$ is zero and $i$ and $j$ are distinguishable.

According to equation 24, it is easily verified that the individuals most uniformly distributed on the property scale have the highest diversity (the first row in Table 1, $M_a = N = 3$, $s_a = 1$ and $S = 0$), where all species are distinguishable, in contrast to a collection of samples as shown in the last row in Table 1, which has the lowest diversity and the highest indistinguishability ($M_a = 1$, $s_a = N = 3$ and $S = 3.29$). For the latter case (the last row in Table 1), because the locations on the property scale of these three individuals are the same, they are the same species and all can be represented by one same symbol, to say B; the 27 indistinguishable microstates are those listed in Figure 5.

The calculation results of several representative sets of samples by using eq 24 for pairwise similarity calculation are listed in Table 1.

**Table 1.** Calculation of diversity based on the similarity formula eq. 24. Eqs 14, 19, 20, 21, 5 and 4 are used for calculating entropy ($S$), apparent indistinguishability number of microstates ($w_a$), apparent indistinguishability number of species or apparent indistinguishability number of symbols ($s_a$), the apparent number of species ($M_a$), information ($I$) and diversity index ($D$), respectively.

| Sample Properties | Pairwise Similarity Table | Probability Table | $S$ | $w_a$ | $s_a$ | $M_a$ | $I$ | $D$ |
|---|---|---|---|---|---|---|---|---|
| A   B   C<br>0.00  0.50  1.00 | 1 0 0<br>0 1 0<br>0 0 1 | 1 0 0<br>0 1 0<br>0 0 1 | 0.00 | 1.00 | 1.00 | 3.00 | 3.30 | 1.00 |
| A   C B<br>0.00  0.50  1.00 | 1 0 0<br>0 1 1<br>0 1 1 | 1 0 0<br>0 0.5 0.5<br>0 0.5 0.5 | 1.39 | 4.00 | 1.59 | 1.89 | 1.91 | 0.58 |
| A         C B<br>0.00  0.50  1.00 | 1 0 0<br>0 1 1<br>0 1 1 | 1 0 0<br>0 0.5 0.5<br>0 0.5 0.5 | 1.39 | 4.00 | 1.59 | 1.89 | 1.91 | 0.58 |
| A C  B<br>0.00  0.50  1.00 | 1 0 0.5<br>0 1 0.5<br>0.5 0.5 1 | 0.67 0 0.33<br>0 0.67 0.33<br>0.25 0.25 0.50 | 2.31 | 10.1 | 2.16 | 1.39 | 0.99 | 0.30 |
| C B A<br>0.00  0.50  1.00 | 1 1 1<br>1 1 1<br>1 1 1 | 1/3 1/3 1/3<br>1/3 1/3 1/3<br>1/3 1/3 1/3 | 3.30 | 27.0 | 3.00 | 1.00 | 0.00 | 0.00 |

**5. Concluding Remarks**

Traditionally the statistical mechanics concepts (e.g, Shannon's entropy [3] and the parameter of E. T. Jaynes [12] which is similar to temperature) and statistical mechanics expressions were introduced to information theory. This paper has described an application of information theory concepts in diversity assessment by introducing a new relation of entropy-similarity, based on the rejection of Gibes paradox statement, after the new relation of entropy-similarity having been used in many areas of physics and chemistry for structural stability and process spontaneity characterization [5-7]. We can see clearly that the similarity (*Z*) of the species in a system (fauna and flora in an ecosystem or a mixture of molecules in a chemical reaction container) [1]

$$Z = 1 - D \tag{26}$$

increases with the decrease in diversity (*D*). The diversity defined here, even though only a relative scale, is useful for comparison, for monitoring the trend of the evolution and for the decision making. Similarity (*Z*), called relative entropy by Shannon, is defined as

$$Z = \frac{S}{S_{max}} \tag{26}$$

which depends directly on the pair-wise similarities.

In contrast to the using of statistical mechanics concepts to information theory, which seems successful so far, the application of information theory to statistical mechanics has been unsuccessful. For instance, Jaynes' theory [12] has only been used to pure data reduction. In order to do this, it is necessary to define clearly the basic concepts. We have demonstrated that it is possible to avoid the widespread and continuous conceptual confusion between the information loss of *dynamic* mixing and the information loss in *static* assembling and to recognize that both satisfy the definition of entropy is information loss (Equation 5).

The application of statistical mechanics already has some foundational problems. An obvious problem is the Gibbs paradox, which is closely related to information theory [8]. The new relation (higher entropy-higher similarity relation, Figure 1c) contrasting to the so far widely accepted relations (Figure 1a or 1b), must have theoretical consequences in many areas. Therefore, the establishment of this new entropy-similarity relation, which is true in both dynamic and static cases, will be a significant step towards the direct application of some methods of information theory in statistical mechanics and thermodynamics. In a following paper, we will use this entropy-information loss definition and the new, higher entropy-higher similarity relation to discuss the symmetry problem [13,14]. A system of very high symmetry must have high entropy value and less information than a less symmetric system. A system of high diversity, which has been discussed here, must appear less symmetric.

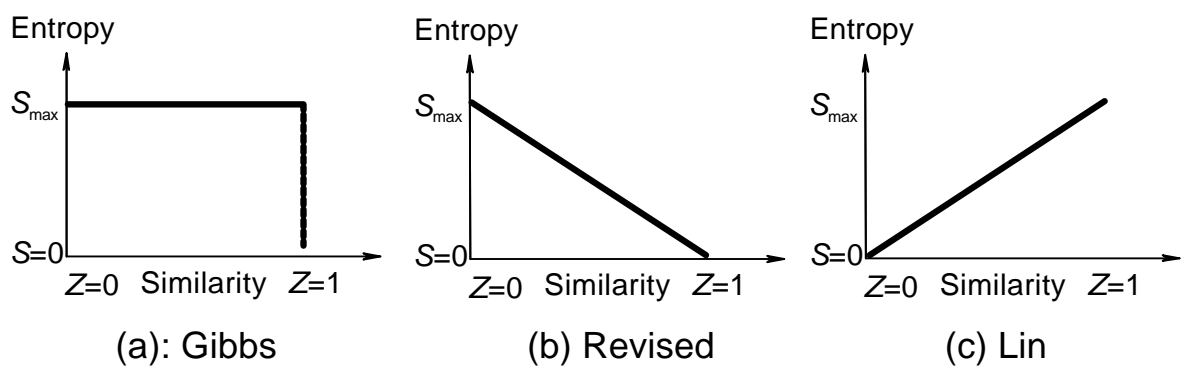

**Figure 1.** Correlation of entropy of mixing (fluid phase) or assembling (static or solid phase) with similarity. Entropy decreases discontinuously with the similarits of the components (Figure a) [8]; decreases continuously (Figure b) [9]; increase continuously (Figure c) [5-7].

# ABC

(a)

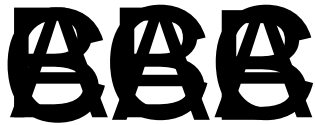

(b)

Figure 2. Schematic representation of information loss due to dynamic motion. The pictures on the three positions are the same, hence they are symmetric. All the three letters appear at any position the same probability at Figure 2b.

# ABC

(a)

# DDD

(b)

Figure 3. Schematic representation of information loss due to inherent similarities. If all the three letters used for information registration are actually look the same (have same properties as D), we have information loss due to the reduced number of species used for information recording or due to the reduction of diversity [1]. Any system of high diversity must have low similarity [1].

00  11

01  10

**Figure 4.** A binary system of distinguishable species ($M_a = 2$) with $N=2$ which gives 4 distinguishable microstates.

```
AAA   BBB   CCC
AAB   BAB   CAA
AAC   BAA   CAC
ABA   BBA   CBB
ABB   BBC   CBC
ACA   BCB   CCA
ACC   BCC   CCB
ABC   BAC   CBA
ACB   BCA   CAB
```

**Figure 5.** A trinary system of distinguishable species ($M=3$) with ($N=3$, $M=3$) which gives 27 distinguishable microstates.

00 00

00 00

**Figure 6.** A binary system with *N*=2 which gives 4 indistinguishable microstates with the highest species indistinguishability ($s_a = 2$).

BBB BBB BBB
BBB BBB BBB
BBB BBB BBB
BBB BBB BBB
BBB BBB BBB
BBB BBB BBB
BBB BBB BBB
BBB BBB BBB
BBB BBB BBB

**Figure 7.** A trinary system with (*N*=3) which gives 27 indistinguishable microstates because of the highest species indistinguishability ($s_a = 3$). An example is shown in Figure 6e, where the property is represented by the symbol "B".